\documentclass[iop]{emulateapj}

\begin{document}

\title{Discovery of Super-Li Rich Red Giants in Dwarf Spheroidal
  Galaxies\altaffilmark{*}}

\slugcomment{Accepted to ApJL on 3 May 2012}

\author{Evan~N.~Kirby\altaffilmark{1,2}, Xiaoting~Fu\altaffilmark{3},
  Puragra~Guhathakurta\altaffilmark{4}, and
  Licai~Deng\altaffilmark{3}}

\altaffiltext{*}{The data presented herein were obtained at the
  W.M. Keck Observatory, which is operated as a scientific partnership
  among the California Institute of Technology, the University of
  California and the National Aeronautics and Space
  Administration. The Observatory was made possible by the generous
  financial support of the W.M. Keck Foundation.}
\altaffiltext{1}{California Institute of Technology, 1200 E.\ California Blvd.,
  MC 249-17, Pasadena, CA 91125, USA}
\altaffiltext{2}{Hubble Fellow}
\altaffiltext{3}{Key Laboratory of Optical Astronomy, National Astronomical
  Observatories, Chinese Academy of Sciences, Beijing, 100012, China}
\altaffiltext{4}{University of California Observatories/Lick Observatory,
  University of California, 1156 High St., Santa Cruz, CA 95064, USA}

\keywords{stars: abundances --- galaxies: dwarf --- Local Group}

\begin{abstract}

Stars destroy lithium (Li) in their normal evolution.  The convective
envelopes of evolved red giants reach temperatures of millions of K,
hot enough for the $^7{\rm Li}(p,\alpha)^4{\rm He}$ reaction to burn
Li efficiently.  Only about 1\% of first-ascent red giants more
luminous than the luminosity function bump in the red giant branch
exhibit $A({\rm Li}) > 1.5$.  Nonetheless, Li-rich red giants do
exist.  We present 15 Li-rich red giants---14 of which are new
discoveries---among a sample of 2054 red giants in Milky Way dwarf
satellite galaxies.  Our sample more than doubles the number of
low-mass, metal-poor (${\rm [Fe/H]} \la -0.7$) Li-rich red giants, and
it includes the most-metal poor Li-enhanced star known (${\rm [Fe/H]}
= -2.82$, $A({\rm Li})_{\rm NLTE} = 3.15$).  Because most of these
stars have Li abundances larger than the universe's primordial value,
the Li in these stars must have been created rather than saved from
destruction.  These Li-rich stars appear like other stars in the same
galaxies in every measurable regard other than Li abundance.  We
consider the possibility that Li enrichment is a universal phase of
evolution that affects all stars, and it seems rare only because it is
brief.

\end{abstract}

\section{Introduction}

Before the first star was born, Li was the third most abundant element
in the universe.  In the intervening 14~Gyr between then and now, Li
has been both created by spallation of carbon, nitrogen, and oxygen
nuclei and destroyed by astration in the interiors of stars.  In most
stars, destruction rates exceeded creation rates.  Because stars
produce rather than destroy most other elements, Li today is among the
least abundant of the elements lighter than Zn.

\addtocounter{footnote}{-1}

In most metal-poor stars, the abundance of Li is a predictable
function of surface temperature.  Spectroscopy of large samples of
stars in the Milky Way's halo \citep{spi82,gra00} and in metal-poor
globular clusters \citep{lin09b,muc11} show that Li abundances in
dwarf stars remain at the same value ($A({\rm Li}) =
2.3$)\footnote{$A({\rm Li}) = 12 + \log [n({\rm Li})/n({\rm H})]$
  where $n$ is the number density of atoms.} until the first dredge-up
on the subgiant branch.  In this surface convection episode, material
from deeper, hotter layers of the stars mixes with material at the
stellar surface.  The deeper layers contain no Li because Li burning
is very efficient at $T \ga 2.5 \times 10^6$~K, cool compared to
hydrogen burning temperatures.  As a result, the first dredge-up
depletes the photospheric value of Li by a factor of 15--20
\citep{lin09b}.

Another dilution episode occurs at the luminosity function bump in the
red giant branch (RGB)\@.  Extra mixing processes
\citep[e.g.,][]{pal11a} further introduce Li-depleted material to the
stellar surface.  The abundance of Li in red giants drops drastically
as the star evolves beyond the RGB bump.  After both dilution
episodes, the number density of Li atoms drops to below 10 parts per
trillion.  Almost all first-ascent red giants with luminosities
greater than the RGB bump have $A(\rm{Li}) < 1.5$.

Some stars exhibit glaring exceptions to the standard picture of Li
evolution.  For example, excess Li often accompanies $^{13}$C
enhancement in CJ stars \citep{hat03}.  These stars could have
participated in ``hot bottom burning'' \citep[a phrase coined
  by][]{sca75}, wherein $^7$Li can be synthesized in the star and
observed at its surface \citep{cam55}.  Specifically, the reaction
$^3{\rm He}(\alpha,\gamma)^7{\rm Be}$, part of the pp-II hydrogen
burning chain, occurs at temperatures greater than $10^7$~K\@.  Li can
be produced from Be by electron capture: $^7{\rm Be}(e^- \nu)^7{\rm
  Li}$.  However, the second reaction must occur at $T < 2.5 \times
10^6$~K, or else the Li will be destroyed by proton capture.  The
proposed Cameron-Fowler (\citeyear{cam71}) mechanism solves the
temperature discrepancy by theorizing that $^7$Be can be brought to
the surface of the star, where it may capture an electron to create
$^7$Li.  The stellar surface is cool enough to preserve Li.  However,
the ongoing convection guarantees that the surface Li atoms do not
last long.  They quickly return to destructive temperatures.  Thus,
the surface composition of Li is a balance between its creation by the
Cameron-Fowler mechanism and its destruction by convection.

Hot bottom burning is effective at producing $^7$Li in asymptotic
giant branch (AGB) stars with masses of about 4--7~$M_{\sun}$
\citep{ibe75,sac92}.  The convective envelopes of these stars reach
layers where $^7$Be is created.  However, some less massive giants on
both the RGB and AGB have been found to be Li-rich
\citep[e.g.,][]{kra99,pal11a,ruc11}.  The convective envelopes of
these stars do not reach layers with $^7$Be.  Therefore, $^7$Be should
not be transported to the surfaces of these stars in the context of
the standard model of stellar evolution.  If the Cameron-Fowler
mechanism is operating in these stars, then it requires ``extra
mixing'' or ``cool bottom processing'' \citep{boo95,sac99} to connect
the base of the convective envelope to deeper regions of the star that
contain Be.  At one time, thermohaline convection was considered as a
source of the extra mixing \citep{cha07}, but the diffusion was later
shown to be too slow to account for the photospheric compositions of
red giants \citep{den11,pal11b}.  Alternative mixing processes include
magnetic buoyancy \citep{bus07,nor08} and rotation \citep{cha10}.

Even though low-mass, Li-rich giants are rare, their existence
challenges the standard theory of stellar evolution.  They have
spawned numerous modifications to the standard model.  Different
explanations depend on the evolutionary state of the star: the RGB
bump \citep{cha00}, the AGB \citep{nol03}, or even anywhere along the
RGB \citep{sac99}.  Furthermore, the composition of the star also
influences the strength of extra mixing and therefore the surface
abundance of Li \citep{sac99}.

Globular clusters and dwarf spheroidal galaxies (dSphs) are excellent
places to search for Li-rich giants.  First, they offer space
densities high enough for efficient observations with multi-object
spectrographs.  Second, the stars are at a uniform distance, which
eases the determination of evolutionary state and Li abundance.

\section{Lithium Measurements}

\begin{figure}
\centering
\includegraphics[width=\linewidth]{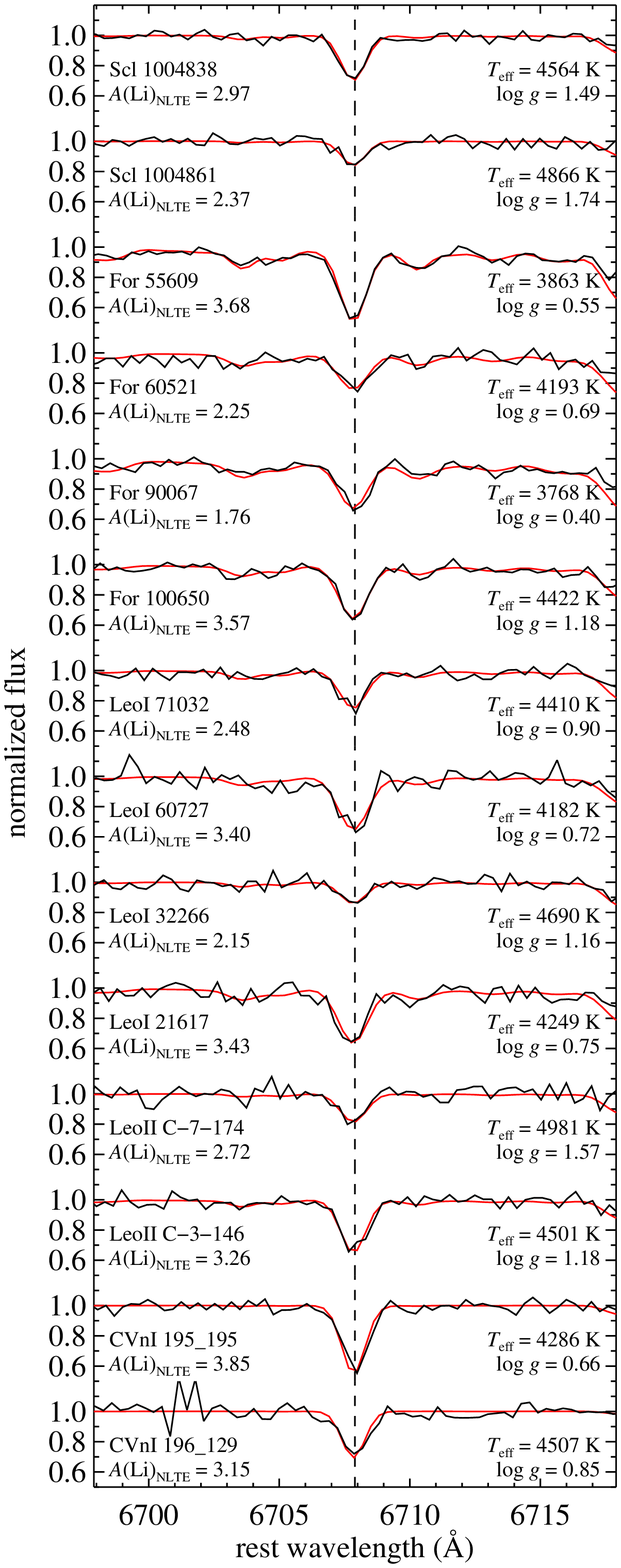}
\caption{Small region of DEIMOS spectra centered on the
  \ion{Li}{1}~6708 multiplet (dashed vertical line) for each of the 14
  dSph giants with detectable Li.  The observed spectra (black) have
  been normalized to have unit continuum.  The red curves show the
  best-fitting synthetic spectra.\label{fig:spectra}}
\end{figure}

\citet{kir10} obtained spectra of nearly 3000 red giants in eight
dSphs with the DEIMOS medium-resolution, multi-object spectrograph
\citep{fab03} on the Keck~II telescope.  Of these data, 2812 spectra
included the spectral region around the \ion{Li}{1} resonance line at
6708~\AA\@.  The slit placement of the other stars caused their
spectra to terminate redward of 6708~\AA\@.  We quantified the
signal-to-noise ratios (S/Ns) of the spectra in the vicinity of the Li
line by computing the inverse standard deviation of
continuum-normalized pixels within 8~\AA\ of the Li line but excluding
the 4~\AA\ immediately surrounding the line.

We searched for detections of the Li line in the 2054 spectra with
${\rm S/N} > 10~{\rm pixel}^{-1}$ that included the appropriate
spectral range.  We found 15 spectra with strong Li lines.  The sample
is random because the stars were not chosen for any property that
could predict Li enhancement.  One of these stars, star 461 in the
Draco dSph, was already known to be Li-rich \citep{dom04}.  The other
stars belong to five dwarf galaxies: Sculptor, Fornax, Leo~I, Leo~II,
and Canes Venatici~I\@.  This sample more than doubles the number of
known Li-rich, metal-poor (${\rm [Fe/H]} \la -0.7$) red giants.

Table~\ref{tab:sample} gives the identities of the 14 newly
discovered, Li-rich stars.  Because the stars reside in different
galaxies, the photometry is not homogeneous.  Table~\ref{tab:sample}
gives the filter set in which each star was observed.
Figure~\ref{fig:spectra} shows the Li-rich stars' spectra around the
Li line, and Table~\ref{tab:abundances} gives the previously measured
\citep{kir10} temperatures, surface gravities, microturbulent
velocities, and metallicities for these stars.

We measured the equivalent widths (EWs) of the Li resonance lines by
fitting Gaussians.  In order to estimate the uncertainty on EW, we
resampled the spectra 1000 times.  In each realization, we perturbed
the flux value of each pixel.  The amount of perturbation was sampled
from a Gaussian random distribution with a width equal to the
measurement uncertainty of the pixel's flux.  The EWs of the detected
Li lines range from 175 to 694~m\AA.  Table~\ref{tab:abundances} gives
the EWs.  We measured the EWs only to illustrate that these lines are
very strong and easily detected.  We used spectral synthesis, not the
EWs, to quantify the Li abundances.

\begin{figure*}
\centering
\includegraphics[width=\linewidth]{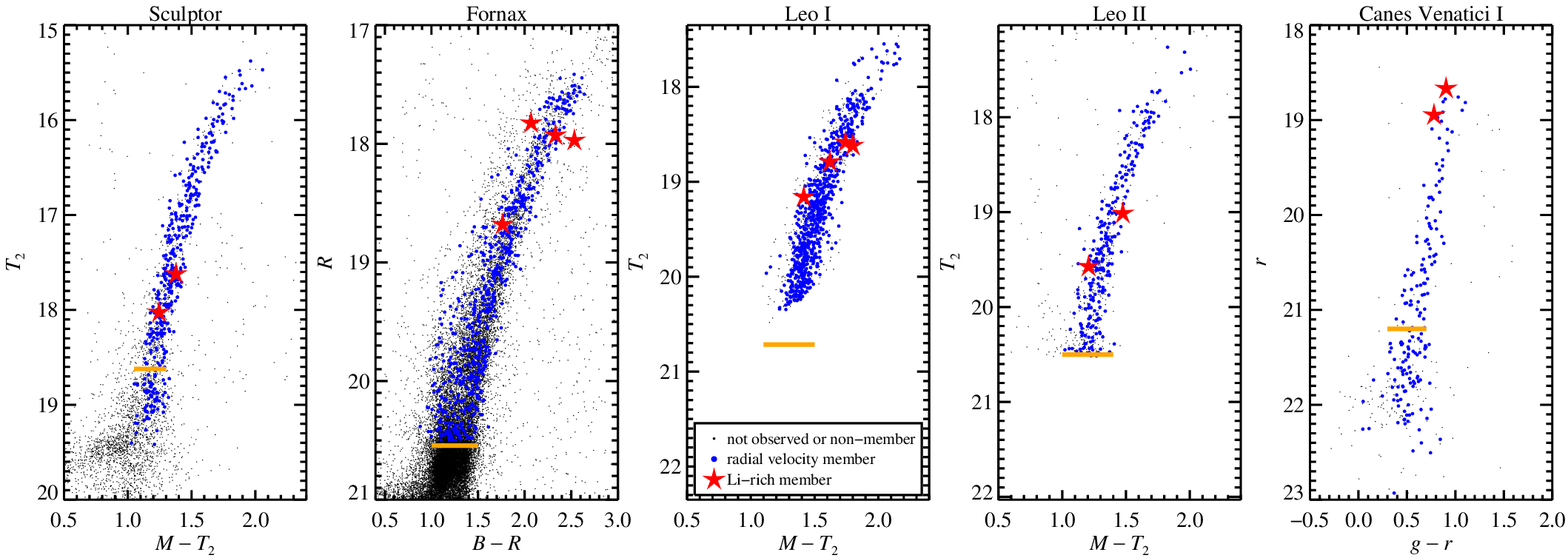}
\caption{Color-magnitude diagrams for the dwarf galaxies in which
  Li-rich giants were detected.  Blue points indicate radial velocity
  members.  Red, five-pointed stars indicate the Li-rich stars, which
  are also radial velocity members.  The orange horizontal lines
  indicate the approximate magnitudes of the RGB bumps.  This
  magnitude was calculated with \citeauthor{fer99}'s
  (\citeyear{fer99}) formula assuming the average age \citep{orb08}
  and metallicity \citep{kir11} of the galaxy.  The colors and
  magnitudes of the Li-rich stars indicate that they are low-mass
  giants more luminous than the RGB bump.\label{fig:cmds}}
\end{figure*}

The Li-rich stars' positions in color-magnitude diagrams (CMDs,
Figure~\ref{fig:cmds}) are consistent with either the RGB or AGB\@.
The colors and magnitudes of the two branches are hardly different for
the old populations typical of dSphs.  Whether the stars in our sample
belong to the RGB or AGB, they belong to dwarf galaxies that are too
old to host the 4--7~$M_{\odot}$ AGB stars that can produce Li in the
standard model of stellar evolution \citep{sac92}.  The colors of the
Li-rich stars are also much redder than intermediate-mass AGB stars.
Therefore, these low-mass stars are anomalous regardless of their
evolutionary states.

\begin{deluxetable*}{lcccccccc}
\tablecolumns{9}
\tablewidth{0pt}
\tablecaption{Li-Rich Red Giant Sample\label{tab:sample}}
\tablehead{\colhead{Star Name} & \colhead{RA (J2000)} & \colhead{Dec (J2000)} & \colhead{Filter 1} & \colhead{Mag 1} & \colhead{Filter 2} & \colhead{Mag 2} & \colhead{Approx. $V$} & \colhead{S/N at 6708~\AA\ (pixel$^{-1}$)}}
\startdata
Scl~1004838          & $00^{\mathrm{h}} 59^{\mathrm{m}} 34 \fs 1$ & $-33 \arcdeg 43 \arcmin 51 \arcsec$ & $  M$ & 19.000 & $T_2$ & 17.622 & 18.746 & 44 \\
Scl~1004861          & $00^{\mathrm{h}} 59^{\mathrm{m}} 34 \fs 2$ & $-33 \arcdeg 43 \arcmin 19 \arcsec$ & $  M$ & 19.272 & $T_2$ & 18.026 & 19.045 & 45 \\
For~55609            & $02^{\mathrm{h}} 39^{\mathrm{m}} 47 \fs 9$ & $-34 \arcdeg 27 \arcmin 47 \arcsec$ & $  B$ & 20.260 & $  R$ & 17.928 & 18.887 & 38 \\
For~60521            & $02^{\mathrm{h}} 39^{\mathrm{m}} 52 \fs 0$ & $-34 \arcdeg 36 \arcmin 31 \arcsec$ & $  B$ & 19.895 & $  R$ & 17.826 & 18.434 & 29 \\
For~90067            & $02^{\mathrm{h}} 40^{\mathrm{m}} 19 \fs 6$ & $-34 \arcdeg 33 \arcmin 42 \arcsec$ & $  B$ & 20.506 & $  R$ & 17.971 & 18.969 & 50 \\
For~100650           & $02^{\mathrm{h}} 40^{\mathrm{m}} 31 \fs 3$ & $-34 \arcdeg 28 \arcmin 52 \arcsec$ & $  B$ & 20.444 & $  R$ & 18.681 & 19.185 & 31 \\
LeoI~71032           & $10^{\mathrm{h}} 08^{\mathrm{m}} 17 \fs 6$ & $+12 \arcdeg 18 \arcmin 19 \arcsec$ & $  M$ & 20.412 & $T_2$ & 18.791 & 19.971 & 34 \\
LeoI~60727           & $10^{\mathrm{h}} 08^{\mathrm{m}} 18 \fs 0$ & $+12 \arcdeg 20 \arcmin 59 \arcsec$ & $  M$ & 20.420 & $T_2$ & 18.619 & 19.941 & 18 \\
LeoI~32266           & $10^{\mathrm{h}} 08^{\mathrm{m}} 30 \fs 1$ & $+12 \arcdeg 17 \arcmin 01 \arcsec$ & $  M$ & 20.575 & $T_2$ & 19.160 & 20.174 & 28 \\
LeoI~21617           & $10^{\mathrm{h}} 08^{\mathrm{m}} 37 \fs 3$ & $+12 \arcdeg 20 \arcmin 12 \arcsec$ & $  M$ & 20.326 & $T_2$ & 18.584 & 19.856 & 21 \\
LeoII~C-7-174        & $11^{\mathrm{h}} 13^{\mathrm{m}} 19 \fs 0$ & $+22 \arcdeg 06 \arcmin 45 \arcsec$ & $  M$ & 20.779 & $T_2$ & 19.573 & 20.477 & 18 \\
LeoII~C-3-146        & $11^{\mathrm{h}} 13^{\mathrm{m}} 36 \fs 2$ & $+22 \arcdeg 08 \arcmin 51 \arcsec$ & $  M$ & 20.489 & $T_2$ & 19.014 & 20.134 & 41 \\
CVnI~195\_195        & $13^{\mathrm{h}} 28^{\mathrm{m}} 27 \fs 6$ & $+33 \arcdeg 36 \arcmin 43 \arcsec$ & $  g$ & 19.571 & $  r$ & 18.667 & 19.044 & 37 \\
CVnI~196\_129        & $13^{\mathrm{h}} 28^{\mathrm{m}} 44 \fs 3$ & $+33 \arcdeg 34 \arcmin 12 \arcsec$ & $  g$ & 19.726 & $  r$ & 18.947 & 19.251 & 26 \\
\enddata
\tablerefs{Identifications and photometry are from \citet{wes06} for Sculptor, \citet{ste98} for Fornax, \citet{soh07} for Leo~I and Leo~II, and the Sloan Digital Sky Survey \citep{ade07} for Canes Venatici~I.}
\end{deluxetable*}

\begin{deluxetable*}{lccccccccc}
\tablecolumns{10}
\tablewidth{0pt}
\tablecaption{Stellar Parameters and Lithium Abundances\label{tab:abundances}}
\tablehead{\colhead{Star Name} & \colhead{$T_{\rm eff}$~(K)} & \colhead{$\log g$~(cm~s$^{-2}$)} & \colhead{$\xi$~(km~s$^{-1}$)} & \colhead{[Fe/H]} & \colhead{EW(\ion{Li}{1}~6708)} & \colhead{$A({\rm Li})_{\rm LTE}$} & \colhead{$A({\rm Li})_{\rm NLTE}$} & \colhead{$\sigma_{\rm noise}$} & \colhead{$\sigma_{T_{\rm eff}}$}}
\startdata
Scl~1004838          & 4564 & 1.49 & 1.79 & $-1.59 \pm 0.11$ & $363 \pm 19$ & $3.32$ & $2.97$ & $0.12$ & $0.24$ \\
Scl~1004861          & 4866 & 1.74 & 1.73 & $-1.70 \pm 0.12$ & $193 \pm 29$ & $2.46$ & $2.37$ & $0.15$ & $0.15$ \\
For~55609            & 3863 & 0.55 & 2.01 & $-0.73 \pm 0.11$ & $694 \pm 24$ & $3.69$ & $3.68$ & $0.10$ & $0.14$ \\
For~60521            & 4193 & 0.69 & 1.98 & $-0.86 \pm 0.11$ & $382 \pm 35$ & $2.11$ & $2.25$ & $0.13$ & $0.20$ \\
For~90067            & 3768 & 0.40 & 2.05 & $-0.68 \pm 0.11$ & $503 \pm 23$ & $2.02$ & $1.76$ & $0.30$ & $0.11$ \\
For~100650           & 4422 & 1.18 & 1.86 & $-0.95 \pm 0.11$ & $492 \pm 28$ & $3.74$ & $3.57$ & $0.14$ & $0.22$ \\
LeoI~71032           & 4410 & 0.90 & 1.93 & $-1.29 \pm 0.11$ & $322 \pm 27$ & $2.50$ & $2.48$ & $0.15$ & $0.23$ \\
LeoI~60727           & 4182 & 0.72 & 1.97 & $-1.42 \pm 0.12$ & $514 \pm 47$ & $3.49$ & $3.40$ & $0.39$ & $0.19$ \\
LeoI~32266           & 4690 & 1.16 & 1.87 & $-1.35 \pm 0.12$ & $175 \pm 30$ & $2.07$ & $2.15$ & $0.13$ & $0.17$ \\
LeoI~21617           & 4249 & 0.75 & 1.97 & $-1.10 \pm 0.11$ & $546 \pm 52$ & $3.53$ & $3.43$ & $0.37$ & $0.19$ \\
LeoII~C-7-174        & 4981 & 1.57 & 1.77 & $-1.24 \pm 0.12$ & $225 \pm 42$ & $2.92$ & $2.72$ & $0.15$ & $0.16$ \\
LeoII~C-3-146        & 4501 & 1.18 & 1.86 & $-1.40 \pm 0.11$ & $449 \pm 30$ & $3.52$ & $3.26$ & $0.16$ & $0.33$ \\
CVnI~195\_195        & 4286 & 0.66 & 1.99 & $-2.61 \pm 0.12$ & $527 \pm 17$ & $3.98$ & $3.85$ & $0.20$ & $0.28$ \\
CVnI~196\_129        & 4507 & 0.85 & 1.94 & $-2.82 \pm 0.13$ & $380 \pm 36$ & $3.64$ & $3.15$ & $0.23$ & $0.27$ \\
\enddata
\end{deluxetable*}

We prepared each spectrum by normalizing to the continuum.  We divided
the observed spectrum by the best-fitting synthetic spectrum
determined by \citet{kir10}.  We fit a B-spline with a breakpoint
spacing of 25~\AA\ to the quotient spectrum, excluding the region
between 6705.9~\AA\ and 6709.9~\AA\@.  This exclusion made the
continuum determination insensitive to the strength of the Li
resonance line.  We divided the observed spectrum by the spline.

We synthesized the spectral region around the Li resonance line with
the spectral synthesis code MOOG \citep{sne73} coupled with ATLAS9
model atmospheres \citep{kur93,kir11pasp} in local thermodynamic
equilibrium (LTE).  We calculated the surface gravity of each star
based on its position in the CMD and interpolation in model isochrones
\citep{dem04}.  The temperatures were based on a combination of
photometry and spectroscopy \citep{kir10}.  The spectra were
synthesized with the multiplet of $^7$Li lines \citep{hob99}.  We
assumed that all of the Li was in the $^7$Li isotope, as is typical
for metal-poor stars \citep{asp06}.  The full spectral range
synthesized was 6697.9--6717.9~\AA.  Although Li is by far the
strongest line in a 5~\AA\ window around 6708~\AA, we supplemented the
line list with atomic (including \ion{Fe}{1}~6707) and molecular (CN,
C$_2$ and MgH) transitions from elements other than Li.  We adopted
the same line list as \citet{kir08}.

We calculated Li abundances and their uncertainties by minimizing
$\chi^2$ (Equation~\ref{eq:chisq}) between the observed and synthetic
spectra using Levenberg-Marquardt optimization.

\begin{equation}
\chi^2 = \sum_{\lambda = 6705.9~{\rm \AA}}^{6709.9~{\rm \AA}} \frac{(f(\lambda) - s(\lambda))^2}{\sigma(\lambda)^2} \label{eq:chisq}
\end{equation}

\noindent In Equation~\ref{eq:chisq}, $f$ represents the
continuum-normalized observed spectrum, $s$ represents the synthetic
spectrum, and $\sigma^2$ represents the variance of the observed
spectrum, propagated through flat fielding and continuum
normalization.  We repeatedly computed synthetic spectra with varying
Li abundances until the $\chi^2$ reached a minimum and changed by less
than one part in $10^5$ between iterations.  Figure~\ref{fig:spectra}
shows the best-fitting synthetic spectra in red.  We applied
corrections for deviations from LTE using \citeauthor{lin09a}'s
(\citeyear{lin09a}) grid.  In some cases, we extrapolated beyond the
boundaries of the grid ($\log g < 1$ and $T_{\rm eff } < 4000$~K).
Table~\ref{tab:abundances} lists the LTE and non-LTE (NLTE) Li
abundances for the 14 newly discovered Li-rich giants.

We calculated two sources of error: $\sigma_{\rm noise}$ and
$\sigma_{T_{\rm eff}}$.  The random error on $A({\rm Li})$ from
spectral noise, $\sigma_{\rm noise}$, is the amount by which $A({\rm
  Li})$ can change before $\chi^2$ increases by one.  There is also
some systematic error from the uncertain effective temperature of the
star, the continuum placement, and uncertainties in the transition
probabilities.  The uncertainty on $T_{\rm eff}$ dominates the
systematic error.  The approximate uncertainty on $T_{\rm eff}$ for
these relatively cool giants is 100~K\@.  We determined this
systematic error by recalculating the best-fitting value of $A({\rm
  Li})$ from synthetic spectra with $T_{\rm eff}$ both 100~K above and
100~K below the nominal value of $T_{\rm eff}$ determined by
\citet{kir10}.  We then corrected $A({\rm Li})$ for NLTE effects.  The
average of these deviations of $A({\rm Li})$ from the value computed
with the unperturbed temperature is $\sigma_{T_{\rm eff}}$.
Table~\ref{tab:abundances} gives both $\sigma_{\rm noise}$ and
$\sigma_{T_{\rm eff}}$.

\section{Discussion}

\begin{figure}
\centering
\includegraphics[width=\linewidth]{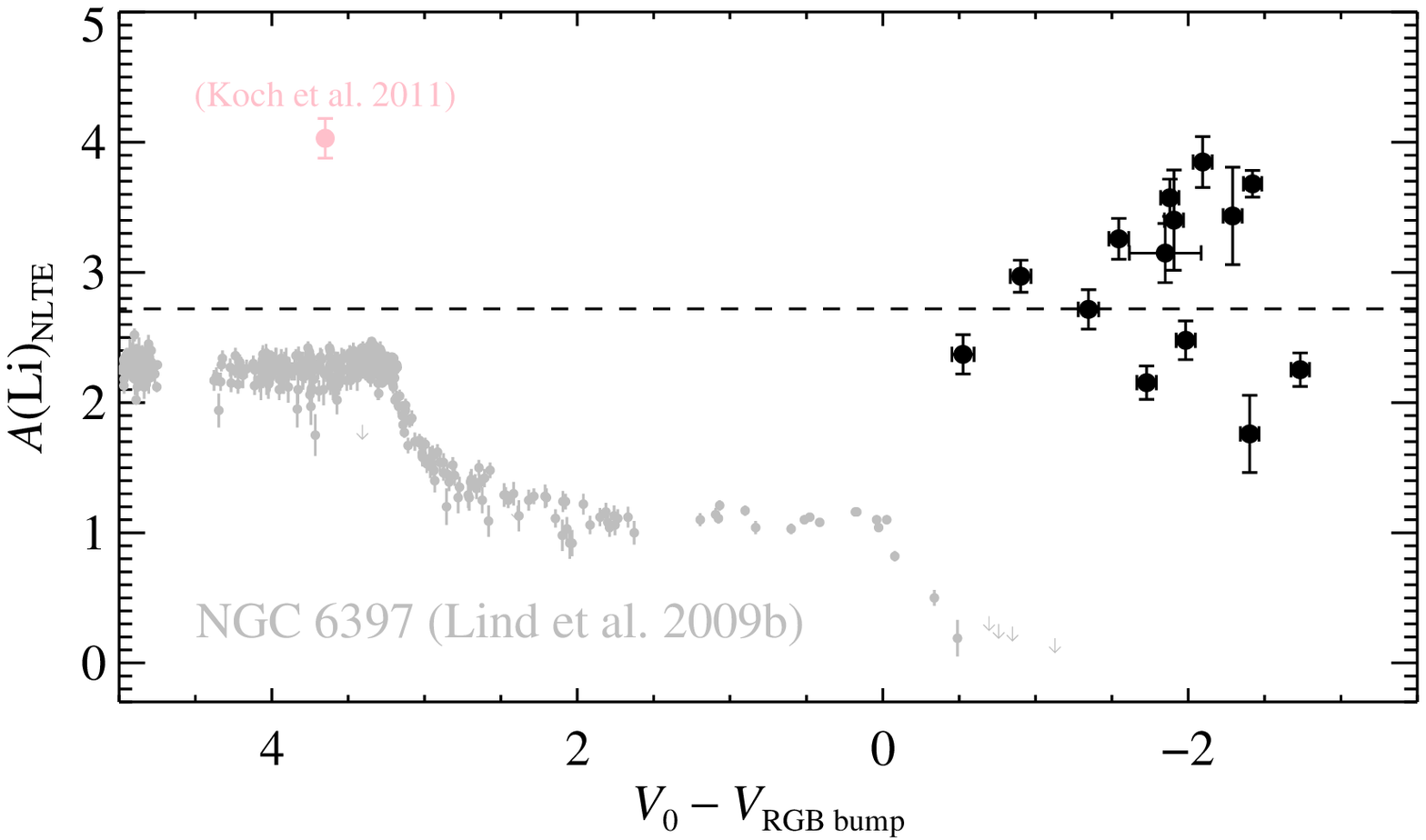}
\caption{Li abundances as a function of the dSph stars' differences in
  $V$ magnitude from the predicted magnitude of the RGB bump,
  calculated with \citeauthor{fer99}'s (\citeyear{fer99}) formula,
  assuming the average age \citep{orb08} of the dSph and the measured
  metallicity of the star \citep{kir10}.  The errors on $A({\rm Li})$
  are given by $\sqrt{\sigma_{\rm noise}^2 + \sigma_{T_{\rm eff}}^2}$.
  Also shown are Li abundances in the globular cluster NGC~6397
  \citep[gray points and upper limits,][]{lin09b}.  The dSph giants in
  our sample are much more Li-enhanced than typical red giants, and
  many are more Li-rich than the universe's primordial Li abundance
  \citep{coc12}, indicated by the dashed line.  NGC~6397 also contains
  an unusually Li-rich turn-off star \citep[pink
    point,][]{koc11}.\label{fig:lin09b}}
\end{figure}

The Li abundances range from $A({\rm Li})_{\rm{NLTE}} = 1.76$ to
$3.85$.  The universe's primordial value of Li is $A({\rm Li}) = 2.72$
\citep{coc12}.  Eight of the stars in our sample have larger Li
abundances.  Therefore, these stars have not merely refrained from
participating in Li destruction.  The Li in these stars must have been
created since the Big Bang.  Our discovery reasserts that the
phenomenon of extreme Li enrichment in giants is not limited to the
Milky Way.  Furthermore, the phenomenon extends to very metal-poor
stars.  The two Li-rich stars in Canes Venatici~I have ${\rm [Fe/H]} <
-2.6$.

Figure~\ref{fig:lin09b} shows $A({\rm Li})$ as a function of the
evolutionary state of the star, expressed as a difference in magnitude
from the RGB bump in the dSph.  The RGB bump magnitude is calculated
individually for each star from \citeauthor{fer99}'s
(\citeyear{fer99}) formula, assuming the mean age of the dSph
\citep{orb08} and the measured metallicity of the star \citep{kir10}.
For comparison, Figure~\ref{fig:lin09b} also includes Li abundances in
the metal-poor globular cluster NGC~6397 from the main sequence
through RGB bump \citep{lin09b}.  Incidentally, NGC~6397 contains a
Li-rich turn-off star, which is even harder to explain than Li-rich
giants \citep{koc11}.

Like most other metal-poor, Li-rich giants \citep{ruc11}, all of the
stars in our sample are more luminous than the RGB bump.  However, our
sample is biased toward high luminosities.  Of the stars we searched,
1764 out of 2054 (86\%) are more luminous than the RGB bump, and the
Li line at fixed abundance becomes weaker for decreasing luminosities
(higher temperatures).  Although our sample does not offer strong
statistical evidence for the Li-rich phenomenon to occur exclusively
above the RGB bump in metal-poor stars, all of the known, metal-poor,
Li-rich giants are consistent with that hypothesis \citep[also
  see][]{gon09}.  Almost all of our sample's stars that are less
luminous than the bump reside in galaxies without detections of
Li-rich stars.

The fraction of strong Li detection in our sample of stars with ${\rm
  S/N} > 10~{\rm pixel}^{-1}$ above the RGB bump is 15 of 1764
(0.85\%).  However, the detectability of Li depends on the stellar
temperature, Li abundance, and spectral S/N.  Our spectra with lower
S/Ns could harbor anomalously large Li lines, but we possibly would
have missed them in our visual search.  Future work (X.~Fu et al., in
preparation) will make a more quantitative determination of the
Li-rich fraction of red giants in our sample.

The existence of Li-rich red giants and their abundances do not
correlate with any measurable parameter.  Although \citet{cha00} found
that Li-rich giants seem to cluster in the CMD, including at the RGB
bump, our sample shows no such clustering.  \citet{mon11} and
\citet{leb12} found a similar result in the Milky Way disk and bulge.
These and our samples have the advantage that the stars are in stellar
systems with known distances.  Therefore, the magnitude distance from
the RGB bump does not need to be inferred from spectroscopically
determined atmospheric parameters.  In addition to positions in the
CMD, our stars' temperatures, surface gravities, iron abundances, and
[$\alpha$/Fe] abundance ratios are not unusual in any regard with
respect to Li-normal stars in the same dwarf galaxies.  Although the
resolution of our spectra is too low to measure rotation, at least
80\% of metal-poor, Li-rich red giants in another survey \citep{ruc11}
exhibit typical rotation velocities.

The typicality of metal-poor, Li-rich stars in all regards except Li
abundance suggests that these stars are not unusual in terms of their
intrinsic properties or external stimuli.  Furthermore, the fraction
of Li-rich, metal-poor giants in our sample shows that the frequency
of the Li-rich phenomenon in dSphs---about 1\%---is roughly the same
as in the Milky Way disk, bulge, and halo
\citep{bro89,ruc11,mon11,leb12}.  The apparent randomness of giants
that exhibit large Li abundances restricts extra mixing models.  For
example, the sudden increase in angular momentum caused by engulfment
of a planet could induce extra mixing \citep{den00}.  However, the
occurrence of planets in the solar neighborhood is far more likely
around metal-rich stars \citep{fis05}.  Assuming that stars in dSphs
also obey a correlation between metallicity and the occurrence of hot
Jupiters and that the occurrence relation extends to very low
metallicities ($-3 < {\rm [Fe/H]} < -0.5$), the stars in our sample
are extremely unlikely to host hot Jupiters.  In fact, we could find
no model in the literature that adequately explains the available
observations for Li-rich, low-mass, metal-poor giants (Li enhancements
as high as $A({\rm Li}) = 3.9$ even near the tip of the RGB, no
concentration in the CMD, weak correlation with rotation).

Because Li-rich, metal-poor giants are otherwise ordinary, we suggest
that Li enhancement does not arise only in special cases.  Instead, we
echo a previous suggestion \citep{del96,gon09} that extra mixing and
its associated Li enhancement could be a brief, universal phase of
stellar evolution.  The lifetime of a 10~Gyr old red giant is 420~Myr,
35~Myr of which is spent above the RGB bump \citep{dot08}.  The rate
of Li depletion with increasing luminosity in NGC~6397 \citep{lin09b}
is roughly $\Delta A({\rm Li}) / \Delta M_V = 1.5$.  Just above the
RGB bump, red giants brighten by one magnitude in 19~Myr
\citep{dot08}.  From these derivatives, we infer that the $e$-folding
time for Li in the atmosphere of a normal, metal-poor red giant near
the RGB bump is about 5~Myr, or 15\% of the lifetime of the red giant
above the RGB bump.  However, the convection zone is deeper closer to
the RGB tip than at the RGB bump, so the Li destruction rate must
accelerate.  Furthermore, the destruction rate could be even faster in
the presence of extra mixing, which brings photospheric material to
even hotter temperatures.  The accelerated Li destruction could
conspire to reduce the observable lifetime of an instantaneously
Li-enhanced star to just 1\% of the lifetime above the RGB bump.  In
this scenario, about 1\% of all red giants above the bump would appear
Li-rich.  \citet{pal01} postulated that this process happens at the
RGB bump in a Li flash that also serves to temporarily increase the
luminosity of the red giant, which could explain why Li-rich giants
are observed at all luminosities between the bump and the tip of the
RGB.  However, neither \citet{den04} nor \citet{pal06} could achieve
high enough mixing rates in their models to trigger a Li flash.
Nonetheless, the idea that Li enhancement is a brief, universal phase
of stellar evolution remains attractive in order to explain the lack
of correlation with almost any other measurable parameter.

\acknowledgments We thank the editor and the anonymous referee for a
timely and helpful report.  Support for this work was provided by NASA
through Hubble Fellowship grant 51256.01 awarded to E.N.K. by the
Space Telescope Science Institute, which is operated by the
Association of Universities for Research in Astronomy, Inc., for NASA,
under contract NAS 5-26555.  X.T.F. and P.G. acknowledge support by
NSF grant AST 09-37525.  X.T.F. and L.D. thank NSFC for support by
grants nos. 10973015 and 11061120454.  PG acknowledges NSF grant
AST-1010039.  The authors wish to recognize and acknowledge the very
significant cultural role and reverence that the summit of Mauna Kea
has always had within the indigenous Hawaiian community.  We are most
fortunate to have the opportunity to conduct observations from this
mountain.

{\it Facility:} \facility{Keck:II (DEIMOS)}

\end{document}